\documentclass[11pt]{article}
\usepackage{amsmath}
\usepackage{graphicx,amssymb,booktabs,multirow}
\usepackage[paperwidth=90mm]{geometry}
\usepackage{anysize}
\usepackage{bm}
\usepackage{natbib,amsmath}
\usepackage{graphicx}
\usepackage{fancyhdr}
\usepackage{rotating}
\marginsize{1in}{1in}{1in}{1in}
\makeatletter
\newcommand*{\indep}{%
	\mathbin{%
		\mathpalette{\@indep}{}%
	}%
}
\newcommand*{\nindep}{%
	\mathbin{
		\mathpalette{\@indep}{\not}
	}%
}
\newcommand*{\@indep}[2]{%
	\sbox0{$#1\perp\m@th$}
	\sbox2{$#1=$}
	\sbox4{$#1\vcenter{}$}
	\rlap{\copy0}
	\dimen@=\dimexpr\ht2-\ht4-.2pt\relax
	\kern\dimen@
	{#2}%
	\kern\dimen@
	\copy0 
}
\makeatother
\usepackage{color}
\usepackage{amsthm}

\makeatother
\newtheorem{theorem}{Theorem}
\newtheorem{lemma}{Lemma}

\newtheorem{assumption}{Assumption}
\newtheorem{corollary}{Corollary}
\usepackage{bbm}

\usepackage{caption}
\usepackage{subfigure}

\def\cov{\textnormal{Cov}}

\newcommand{\R}{{\mathbb{R}}}

\usepackage[all]{xy}
\usepackage{bm}
\usepackage{pifont}
\usepackage{stmaryrd}
\usepackage{pgf,tikz}
\usepackage{tikz}
\usetikzlibrary{shapes,arrows,positioning,calc}

\newenvironment{sequation*}{\begin{equation*}\small}{\end{equation*}}

\newenvironment{tequation*}{\begin{equation*}\tiny}{\end{equation*}}

\usepackage{mathrsfs}

\def\sig{\sigma}

\def\pr{\textnormal{pr}}
\def\calS{\mathcal{S}}
\def\calV{\mathcal{V}}
\graphicspath{ {./figures/} }

\newtheorem{innercustomgeneric}{\customgenericname}
\providecommand{\customgenericname}{}
\newcommand{\newcustomtheorem}[2]{%
	\newenvironment{#1}[1]
	{%
		\renewcommand\customgenericname{#2}%
		\renewcommand\theinnercustomgeneric{##1}%
		\innercustomgeneric
	}
	{\endinnercustomgeneric}
}

\newcustomtheorem{customthm}{Theorem}
\newcustomtheorem{customlemma}{Lemma}
\newcustomtheorem{customassumption}{Assumption}
\tikzstyle{sum} = [draw, circle, minimum size=0pt, inner sep = 1.7pt]

\makeatletter
\newcommand{\mathleft}{\@fleqntrue\@mathmargin0pt}
\newcommand{\mathcenter}{\@fleqnfalse}
\makeatother
\newtheorem{example}{Example}
\newtheorem{definition}{Definition}

\usepackage[ruled,linesnumbered]{algorithm2e}
\usepackage{wrapfig}

\allowdisplaybreaks

\usepackage{accents}

\def\T{{ \mathrm{\scriptscriptstyle T} }}

\newcommand\XtoY{{X\rightarrow Y}}
\newcommand\YtoX{{Y\rightarrow X}}

\def\calH{\mathcal{H}}

{
	\theoremstyle{plain}

}

\DeclareFontFamily{U}{mathx}{}
\DeclareFontShape{U}{mathx}{m}{n}{<-> mathx10}{}
\DeclareSymbolFont{mathx}{U}{mathx}{m}{n}
\DeclareMathAccent{\widecheck}{0}{mathx}{"71}
\def\CI{\textnormal{CI}}

\DeclareMathOperator* {\argmax} {arg\,max}
\makeatletter
\newcommand*{\rom}[1]{\expandafter\@slowromancap\romannumeral #1@}
\makeatother

\def\I{\textnormal{I}}
\def\II{\textnormal{II}}
\newtheorem{assumptionalt}{Assumption}[assumption]

\newenvironment{assumptionp}[1]{
	\renewcommand\theassumptionalt{#1}
	\assumptionalt
}{\endassumptionalt}

\usepackage{apptools}
\AtAppendix{\counterwithin{lemma}{section}}

\makeatletter
\DeclareRobustCommand{\pdot}{\mathbin{\mathpalette\pdot@\relax}}
\newcommand{\pdot@}[2]{%
	\ooalign{%
		$\m@th#1\circ$\cr
		\hidewidth$\m@th#1\cdot$\hidewidth\cr
	}%
}
\makeatother

\usepackage{xr}
\usepackage{fullwidth}

\begin{document}
	\title{Discovery and inference of possibly bi-directional causal relationships with invalid instrumental variables}
	\date{}
\author{Wei Li$^1$, Rui Duan$^2$, and Sai Li$^3$\\
	\\
	$^1$ Center for Applied Statistics and School of Statistics, Renmin University of China\\
	$^2$ Department of Biostatistics, Harvard University\\
	$^3$ Institute of Statistics and Big Data, Renmin University of China}
	\maketitle

	\begin{abstract}
		    Learning causal relationships between pairs of complex traits from observational studies is of great interest across various scientific domains. 
      However, most existing methods assume the absence of unmeasured confounding and  restrict causal relationships between two traits to be uni-directional, which may be violated in  real-world systems. In this paper, we address the challenge of causal discovery and effect inference for two traits while accounting for unmeasured confounding and potential feedback loops. By leveraging possibly invalid instrumental variables, we provide  identification conditions for causal parameters in a model that allows for bi-directional relationships, and we also establish identifiability of the causal direction under the introduced conditions.      
      Then we propose a data-driven procedure to detect the causal direction and provide inference results about causal effects along the identified direction. We show that our method consistently recovers the true direction and produces valid confidence intervals for the causal effect. We conduct extensive simulation studies to show that
      our proposal outperforms existing methods. We finally apply our method to analyze real data sets from UK Biobank.
      
      

	\end{abstract}


\section{Introduction}

\subsection{Motivation}
Understanding causal relationships between pairs of complex traits or systems is of fundamental importance in various scientific domains. A common approach for causal discovery is through experimentation. However, the required experiments can sometimes be expensive, unethical, or even impractical. Consequently, the methods for deducing causality from observational studies have gained increasing attention in recent years \citep{shimizu2006linear,mooij2016distinguishing,bongers2021foundations,xu2022inferring}. Despite significant methodological advances,
the majority of such studies rely on two crucial assumptions, namely, no unmeasured confounding and uni-directional relationship between the two traits.

The absence of unmeasured confounding essentially requires investigators to accurately measure covariates that cover all potential sources of confounding. In practice, confounding mechanisms can rarely be accurately learned  from measured covariates, and  unmeasured confounding is a prevalent issue, posing a great challenge for  causal discovery and effect estimation. While the uni-directional assumption permits a straightforward interpretation of  causality between two traits, it may be restrictive in some practical scenarios. For example, in the classical demand-supply model, demand has an effect on supply and vice versa \citep{hyttinen2012learning}. Because a cause always precedes its effect, the relationship between two variables may be intuitively understood as uni-directional over time, such as the demand of the previous time step affecting the supply of the subsequent time step. However, the measures of demand and supply are typically presented as cumulative averages over much longer intervals, obscuring the faster interactions between these two variables. Similar phenomena are common in many biological and epidemiological problems, where bi-directional causal relationships exist between certain traits, such as the relationship between insomnia and major psychiatric disorders \citep{gao2019bidirectional}. In this article, we focus on causal discovery and  effect estimation for the  case of two traits while simultaneously accounting for  unmeasured confounding and reverse causation.

The instrumental variable (IV) estimation is a widely used approach for mitigating confounding bias, the validity of which requires instruments that are associated with the exposure, independent of unmeasured confounding, and have no direct effect on the outcome. Mendelian randomization (MR), a natural counterpart to randomized experiments, is a typical IV method used in genetic epidemiology \citep{smith2004mendelian}. It leverages genetic variants, often single nucleotide polymorphisms (SNPs),  as instruments. These genetic variants are considered randomly inherited from parents to offspring during conception, and therefore, are not subject to confounding. However, in the presence of bi-directional relationships, it is difficult to specify valid instruments for either direction because genetic variants could have their primary influence on the exposure or the outcome due to reverse causation \citep{davey2014mendelian}. 
Besides, some genetic variants may have horizontal pleiotropic effect when they affect the
outcome through  pathways other than the exposure under investigation, which is also known as violation of the exclusion restriction condition for valid IVs \citep{burgess2017interpreting}. While many methods have been developed to relax the exclusion restriction, 
the majority of them focus on causal effect estimation in the uni-directional case with the direction information given apriori. Few methods have formally studied how to infer causal direction and subsequently make inference on causal effects, especially for possibly bi-directional causal relationships.

\subsection{Prior works}

When the relationship between two traits is uni-directional with known directions, various methods have been proposed to address invalid IVs. One strategy assumes that only a small proportion of instruments are invalid. For instance, \citet{han2008detecting,bowden2016consistent,kang2016instrumental,windmeijer2019use} provided methods to recover causal effects under the majority rule where more than 50\% of the relevant instruments are required to be valid. 
Extending this idea, \citet{hartwig2017robust}  and \citet{guo2018confidence} introduced the plurality rule which assumes that the number of valid IVs is greater
than the largest number of invalid IVs with an equal ratio estimator limit. \citet{guo2018confidence} also developed the method of
two-stage hard thresholding (TSHT) with voting for estimation. Another strand of work allows all IVs to be invalid but makes additional assumptions on the data generating process. \citet{bowden2015mendelian} introduced the MR-Egger regression method to provide consistent estimates under the assumption that the effects of the instruments on the exposure are uncorrelated with the invalid effects. \citet{zhao2019statistics,ye2021debiased} addressed balanced horizontal pleiotropy assuming that the mean of the invalid effects is zero.  Alternatively, \citet{tchetgen2021genius} extended the method proposed by \citet{lewbel2012using}, utilizing a covariance heterogeneity restriction for identification. However, all these methods rely on prior knowledge of the uni-directional relationship with known directions. Specifically, it requires specifying that
one trait is the exposure and the other is the outcome, and the causal relationship, if exists, can be only from the exposure to the outcome.

When there is a uni-directional relationship with unknown directions, several methods apply the commonly-used identification rules, such as the valid rule, majority rule, and plurality rule, for each direction separately  to determine causal directions \citep{hemani2017orienting,xue2020inferring,chen2023discovery}. For example, a recent paper by \cite{chen2023discovery} assumed the existence of valid IVs for each primary variable, and then proposed a majority rule-based causal discovery method for Gaussian directed acyclic graphs with unmeasured confounding, which is also applicable for cases involving only two traits. Our work differs from these existing methods as we introduce a different set of identification conditions, which only requires valid instruments in one direction rather than both directions.


In the presence of bi-directional relationships, bi-directional MR is often applied to infer causal direction between two traits \citep{davey2014mendelian,sanderson2022mendelian}. The common practice is to conduct two MR analyses 	on the same pair of phenotypes by reversing 	the exposure and the outcome, which requires  valid instruments for both variables. \citet{davey2014mendelian} suggested using genetic variants with sufficient understanding of their biological effects, which is unfortunately not always  feasible for many traits of interest, and hence, findings from bi-direction MR studies should be interpreted with caution \citep{davey2014mendelian,sanderson2022mendelian}. Recently, in the presence of potentially invalid instruments, \citet{darrous2021simultaneous} proposed a latent heritable confounder MR method to estimate bi-directional causal effects using maximum likelihood. However, this method imposes
specific assumptions about error distributions and hierarchical priors on model parameters, which may be sensitive and computationally unstable. \citet{xue2022robust} developed a constrained maximum likelihood based approach to infer bi-directional  relationships, assuming that the plurality rule condition holds for both directions. 
\citet{li2022focusing} showed that  the  commonly-used identification conditions cannot simultaneously hold for both directions under bi-directional  models. They further developed a focusing framework to infer bi-directional  relationships  using hypothesis testing, but how to estimate causal effects and construct confidence intervals  under such circumstances still remains unclear.

\subsection{Our contributions}

By incorporating  invalid IVs, this paper proposes a complete solution for causal discovery and inference about possibly bi-directional relationships in the presence of unmeasured confounding between a pair of traits. Our work has the following distinctive features and makes several major contributions to  causal  inference studies. 

	First, we propose two identification approaches for causal parameters and directions within possibly bi-directional models that account for unmeasured confounding. In the first approach,  if the plurality rule holds for one known direction and the errors of the other direction satisfy the covariance heterogeneity  restriction, then we prove that the casual parameters are identifiable in bi-directional models. In the second approach, we study the more practical scenario where we do not know in which direction the plurality rule holds. In this case, the plurality rule and covariance heterogeneity conditions are insufficient for parameter identification, as shown in a counter-example. To deal with this challenge, we slightly strengthen the plurality rule and introduce the so-called enhanced plurality rule. We establish identification results for the causal parameters and directions under the enhanced plurality rule and covariance heterogeneity conditions when it is not known a priori in which direction the enhanced plurality rule is true. To the best of our knowledge, this is the first study to address identification issues simultaneously considering unmeasured confounding and reverse causation.

	Second, we develop a data-dependent method  termed PCH to infer the causal relationship and make inference for the causal effects. The PCH involves a modified TSHT for estimation in one direction and a new method based on the covariance heterogeneity condition for the other direction. The modified TSHT accounts for heteroscedastic errors and calculates variance based on the empirical noises. Besides, by introducing novel steps to detect the failure of the enhanced plurality rule and covariance heterogeneity conditions, our procedure can determine the direction in which the enhanced plurality rule holds.
	We demonstrate that the proposed procedure can consistently recover the causal direction between two traits and produce valid confidence intervals for the pair of causal effects.

\subsection{Organization}

The rest of the paper proceeds as follows. Section~\ref{sec:model} introduces the proposed model that accounts for both bi-directional relationships and unmeasured confounding. Section~\ref{sec:identification}  presents  identification conditions and establishes identifiability of causal parameters and directions. Section~\ref{sec:method} describes the proposed method for causal  discovery, provides confidence intervals for causal effects, and offers theoretical justification about the proposed method. 
We then investigate the finite-sample performance of our method via simulation studies in Section~\ref{sec:simulation}, and apply the method to  real data examples from UK Biobank in Section~\ref{sec-data}. We finally conclude with a discussion in Section~\ref{sec:discussion} and relegate proofs to the supplementary material.

We introduce some notations that will be used in the remainder of this paper. For a set $\mathcal{J}$, we denote its complement as $\mathcal{J}^c$. For any $p\times p$ matrix $M$ and sets $\mathcal{A},\mathcal{B}\subseteq\{1,\ldots,p\}$, let $M_{\mathcal{A},\mathcal{B}}$ represent the submatrix formed by the rows specified by the set $\mathcal{A}$ and the columns specified by the set $\mathcal{B}$. We further define $M_{\mathcal{A}\cdot}=M_{\mathcal{A},\mathcal{B}}$ with $\mathcal{B}=\{1,\ldots,p\}$.	
Given any column vector $K\in\R^n$ and an $n\times m$ matrix $Q$, 
we define the elementwise product $K\odot Q=(K_1Q_{1},\ldots,K_nQ_{n})^\T\in\R^{n\times m}$, where $Q_{i}$ is the $i$-th row of $Q$ 
for $i\in\{1,\ldots,n\}$ and
$m\in\{1,\ldots,p\}$. 

\section{Model set-up}\label{sec:model}

Suppose that there are $n$ independent and identically
distributed observations from a population of interest. For the $i$-th observation, let $X_i\in\R$ and $Y_i\in\R$ denote a pair of continuous 
primary variables, $Z_i\in\R^p$ a vector of 
IVs, which are possibly invalid.	
Without loss of generality, we assume that the variables of each observation have been centered, i.e., $E(X_i)=E(Y_i)=0$, and $E(Z_i)=0_p$.  We define $\Sigma=E(Z_iZ_i^\T)$ and assume that $\Sigma$ is invertable. Additionally, we assume that $X_i$ and $Y_i$ have finite variance.
We further define the following vectors (matrices) to denote a collection of $n$ observations:  $X=(X_1,\cdots,X_n)^\T$, $Y=(Y_1,\cdots,Y_n)^\T$, and $Z=(Z_1,\cdots,Z_n)^\T$.	
We  let $U_i$ denote all unmeasured common causes of $X_i$ and $Y_i$. Following the framework   proposed by \cite{bollen1989structural} and \cite{pearl2009causality}, we consider the linear structural equation model (SEM) with explicit unmeasured confounding:
\begin{equation}\label{eqn:sem}
	\begin{aligned}
		Y_i=& \beta_{X\rightarrow Y}X_i+\pi_Y^\T Z_i+\xi_Y(U_i)+\zeta_{i}, \\
		X_i=&\beta_{Y\rightarrow X}Y_i+\pi_X^\T Z_i+\xi_X(U_i)+\eta_{i},\\
	\end{aligned}
\end{equation}
where $\xi_Y(U_i),\xi_X(U_i)$ are two arbitrary functions of $U_i$ with mean zero, and 
$\zeta_{i},\eta_i$ are  noise terms satisfying $E(\zeta_{i}\mid \eta_i,Z_i,U_i)=E(\eta_{i}\mid \zeta_i, Z_i,U_i)=0$. In model~\eqref{eqn:sem},  the candidate IVs $Z_i$ are assumed to be correlated with the primary variables $(X_i, Y_i)$, and  independent of the unmeasured confounders $U_i$. However, since $\pi_X$ and $\pi_Y$ may not be zero,  $Z_i$ may violate the exclusion restriction assumption, which is needed in the conventional IV methods.

\begin{figure}[h!]
	\includegraphics[width=0.95\textwidth]{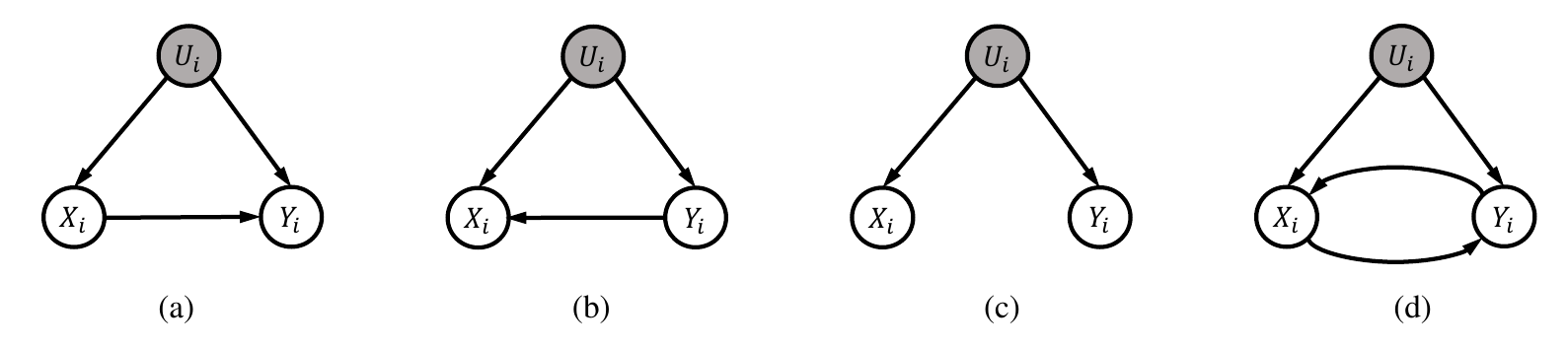}
	\caption{Possible relations of two primary variables $X_i$ and $Y_i$ including  unmeasured confounding $U_i$.
	}\label{fig:dags-model}
\end{figure}

A graphical representation of the SEM in \eqref{eqn:sem} is depicted as Figure \ref{fig:dags-model} where candidate IVs  are omitted. As shown in Figure \ref{fig:dags-model}, there are essentially four different ways how $X_i$ and $Y_i$ can be related  to each
other: (a) $X_i$ causes $Y_i$; (b) $Y_i$ causes $X_i$; (c) $X_i$ and $Y_i$ have no causal connection; (d) $X_i$ and $Y_i$ have cyclic connections. The first three cases (a)-(c) are said to be  {\it uni-directional or acyclic}, which
corresponds to $\beta_\XtoY\beta_\YtoX=0$ in model \eqref{eqn:sem}. Case (d) is said to be {\it bi-directional or cyclic} and corresponds to $\beta_\XtoY\beta_\YtoX\neq 0$ in model \eqref{eqn:sem}.
In
this paper we do not assume a priori that the underlying model is uni-directional. In other words, our model
allows for both bi-directional and uni-directional cases. The focus of this paper is to develop methods that distinguish between the four cases (a)-(d) and make inference on the parameters $(\beta_\XtoY,\beta_\YtoX)$ in model \eqref{eqn:sem}.

In the uni-directional case, the interpretation of the parameters $(\beta_\XtoY,\beta_\YtoX)$ is straightforward. Specifically, if $\beta_\XtoY\neq 0$ and $\beta_\YtoX=0$, then the causal direction is $\XtoY$, as shown in Figure~\ref{fig:dags-model}(a), and $\beta_\XtoY$ is the causal effect of $X_i$ on $Y_i$; if $\beta_\XtoY= 0$ and $\beta_\YtoX\neq 0$, then the causal direction is $\YtoX$, as shown in Figure~\ref{fig:dags-model}(b), and $\beta_\YtoX$ is the causal effect of $Y_i$ on $X_i$; if $\beta_\XtoY=\beta_\YtoX=0$, then there is no causal relation between $X$ and $Y$, as shown in Figure~\ref{fig:dags-model}(c).
The uni-directional case is also known as a recursive SEM in the
literature and corresponds to a causal Bayesian network
with linear relationships over continuous variables \citep{spirtes2001causation}. The interpretation of model (\ref{eqn:sem}) in the bi-directional case, as depicted  in case (d) of Figure \ref{fig:dags-model}, requires more illustrations. The model in this case is often used to represent a causal process that is collapsed over the time dimension and
is typically interpreted as a
dynamical system that equilibrates. More discussions about cyclic models can be found in \citet{koster1996markov,eberhardt2010combining,hyttinen2012learning,heinze2018causal,bongers2021foundations,rothenhausler2021anchor}.

Below we follow \cite{hyttinen2012learning} and \cite{rothenhausler2021anchor} to interpret  model \eqref{eqn:sem} in the bi-directional case. We let $V_i=(Y_i,X_i)^\T$ denote the vector of primary variables, and
$ R_i=(R_{Y,i},R_{X,i})^\T$  the corresponding disturbances, where $R_{Y,i}=\xi_Y(U_i)+\zeta_{i}$ and $R_{X,i}=\xi_X(U_i)+\eta_{i}$. We further define the vector of parameters $\pi=(\pi_Y^{\T},\pi_X^{\T})^{\T}$ and
the  matrix
\begin{align*}
	B=\Bigg(\begin{matrix}
		0 &\beta_{X\rightarrow Y}\\
		\beta_{Y\rightarrow X} & 0
	\end{matrix}
	\Bigg).
\end{align*}
Then the structural equations in \eqref{eqn:sem}  can be rewritten in the following matrix form:
\begin{align*}
	V_i=BV_i+\pi^\T Z_i+R_i.
\end{align*} 
The traditional interpretation of cyclic
SEMs assumes that the exogeneous variables represent background conditions that do not change over the time until the
system has reached equilibrium, which corresponds to what \cite{lauritzen2002chain} called a
deterministic equilibrium. Let $V_i(0)$ denote the initial values of $V_i$ and the data are generated by iterating the system:
\begin{align*}
	V_i(t)=BV_i(t-1)+\pi^\T Z_i+R_i, \quad t\geq 1.
\end{align*} 
Then $\beta_\XtoY$ and $\beta_\YtoX$ are the causal effects of $X_i(t-1)$ on $Y_i(t)$ and $Y_i(t-1)$ on $X_i(t)$, respectively. At time $t$, we have
\begin{align*}
	V_i(t)=B^t V_i(0)+\sum_{k=0}^{t-1}B^k(\pi^\T Z_i+R_i).
\end{align*}
A necessary and sufficient condition for $V_i(t)$ converging to an equilibrium is that the  spectral norm  of $B$ is strictly smaller than one. In that case, we have $B^t\rightarrow 0$ and $\sum_{k=0}^{t-1}B^k\rightarrow (I-B)^{-1}$ as $t$ goes to infinity, and $V_i(t)$ converges to
\begin{align}\label{eqn:reduced-form}
	V_i = (I-B)^{-1}(\pi^\T Z_i+R_i),
\end{align}
where $I-B$ is guranteed to be invertible given the restrictions of the  spectral norm  of $B$. This implies that $\beta_\XtoY\beta_\YtoX\neq 1$. The model in \eqref{eqn:reduced-form} is exactly the reduced-form of the structural equations in \eqref{eqn:sem}.

As assumed, $E(R_{Y,i}\mid Z_i)=E(R_{X,i}\mid Z_i)=0$.
We then obtain 
the following reduced-form equations for the primary variables from \eqref{eqn:reduced-form}: $E(Y_i\mid Z_i)=\gamma_Y^\T Z_i$ and  $E(X_i\mid Z_i)=\gamma_X^\T Z_i$,
where
\begin{align*}
	\gamma_Y=\frac{\pi_Y+\beta_{X\rightarrow Y}\pi_X}{1-\beta_{X\rightarrow Y}\beta_{Y\rightarrow X}},\qquad\text{and}\qquad \gamma_X = \frac{\pi_X +\beta_{Y\rightarrow X}\pi_Y}{1-\beta_{X\rightarrow Y}\beta_{Y\rightarrow X}}.
\end{align*}
According to these two expressions, we have
\begin{align}\label{eqn:ivequation}
	\gamma_Y=\pi_Y+\beta_\XtoY\gamma_X,\quad\text{and}\quad \gamma_X=\pi_X+\beta_\YtoX\gamma_Y.
\end{align}
As $\gamma_Y$ and $\gamma_X$ are identifiable based on the data, equation \eqref{eqn:ivequation} has $2p+2$ unknown parameters but only provides $2p$ equations. Hence, the causal parameters cannot be directly identified without further assumptions.


To determine the parameters $\beta_\XtoY$ and $\beta_\YtoX$, a naive approach would be to separately apply identification rules commonly used in the uni-directional relationships with known directions, such as the valid rule, majority rule, and plurality rule, to the two equations in~\eqref{eqn:ivequation}. For instance, to apply the valid rule, we could assume $\pi_Y=0$, making $\beta_\XtoY$ identifiable. Similarly, assuming $\pi_X=0$ would allow us to identify $\beta_\YtoX$. However, as discussed in \cite{li2022focusing}, these identification rules cannot be simultaneously valid for both directions, and it is impossible to directly apply them to infer bi-directional causal effects. 


\section{Identification of causal effects and directions}
\label{sec:identification}

In this section, we study the identification of causal effects and directions in model (\ref{eqn:sem}).
In Section \ref{subsec:bi-identification}, we develop new identification rules of bi-directional causal effects with possibly invalid IVs when the causal relationship is known to be bi-directional. 
In Section \ref{subsec:uni-identification}, we study the identification of causal effects when the causal relationship is known to be one-directional. In Section \ref{subsec:direction-ident}, we establish the identification of causal directions. These results together provide a complete picture for the identification of causal parameters in (\ref{eqn:sem}) without prior knowledge of the causal directions.

\subsection{Bi-directional causal effect identification}
\label{subsec:bi-identification}

%


We classify all the candidate IVs into four categories. Specifically, let $\mathcal{V}_{null}=\{j:\pi_{X,j}=\pi_{Y,j}=0\}$ denote the set of null IVs, $\mathcal{V}_{\XtoY}=\{j:\pi_{X,j}\neq 0 ,\pi_{Y,j}=0\}$ the set of valid IVs for identifying $\beta_{\XtoY}$, $\mathcal{V}_{\YtoX}=\{j:\pi_{Y,j}\neq 0 ,\pi_{X,j}=0\}$ the set of valid IVs for identifying  $\beta_{\YtoX}$, and  $\mathcal{V}_{pl}=\{j:\pi_{X,j}\neq 0,\pi_{Y,j}\neq 0\}$ the set of pleiotropic IVs. 
Note that our definition of valid IVs agrees with the conventional definition. Indeed, a valid IV for identifying $\beta_{\XtoY}$ is usually defined as having $\gamma_{X,j}\neq 0, \pi_{Y,j}=0$, which according to \eqref{eqn:ivequation} is equivalent to our definition that $\pi_{X,j}\neq 0 ,\pi_{Y,j}=0$.
These four sets $\mathcal{V}_{null}$, $\mathcal{V}_{\XtoY}$, $\mathcal{V}_{\YtoX}$, and $\mathcal{V}_{pl}$ provide a mutually exclusive and exhaustive partition of the set of candidate IVs. 

Let us review the plurality rule proposed by \cite{guo2018confidence}, which is used to deal with invalid IVs in one-directional causal effect models.	Define the set of relevant IVs for $X$ as $\mathcal{S}_X=\{j:\gamma_{X,j}\neq 0\}$ and the set of relevant invalid IVs as $ \mathcal{I}_{\XtoY}=\{j:\gamma_{X,j}\neq 0, \pi_{Y,j}\neq 0\}$. 
The plurality rule  for the direction $X\rightarrow Y$ assumes that
\begin{align}\label{eqn:plurality}
	|\calV_\XtoY|>\max_{c}\Big|\Big\{j\in \mathcal{I}_{\XtoY}:\frac{\pi_{Y,j}}{\gamma_{X,j}}=c\Big\}\Big|.
\end{align}

In the possible existence of bi-directional causal effects, it is easy to see that the relevant invalid IVs for the direction $\XtoY$ are
\[
\mathcal{I}_{\XtoY}=\left\{\begin{array}{ll}
	\calV_{\YtoX} \cup \{\calV_{pl}\cap\calS_{X}\}~&\text{if}~\beta_{\YtoX}\neq 0;  \\
	~\calV_{pl}\cap\calS_{X}~&\text{if}~\beta_{\YtoX}=0.
\end{array}\right.
\]
That is, when the reverse causal effect exists ($\beta_{\YtoX}\neq 0$), the valid IVs for the reverse direction ($\calV_{\YtoX}$) become invalid IVs for the target direction. What makes it worse, the ratios $\{\pi_{X,j}/\gamma_{Y,j}\}_{j\in\calV_{\YtoX}}$ are all equal when $\beta_{\YtoX}\neq 0$. See Lemma 1 of \citet{li2022focusing} for a formal statement. In view of the decomposition of $\mathcal{I}_{\XtoY}$ as above, the plurality rule in \eqref{eqn:plurality} can be rewritten as follows: \begin{align}\label{eqn:pluralityrewrite}
	|\calV_\XtoY|>\max\bigg(\max_{c}\Big|\Big\{j\in\calV_{pl}\cap\calS_X:\frac{\pi_{Y,j}}{\gamma_{X,j}}=c \Big\} \Big|,|\calV_{\YtoX}|\mathbbm{1}(\beta_{\YtoX}\neq 0)\bigg).
\end{align}
The derivation of \eqref{eqn:pluralityrewrite} is given in Section 
S1.1
of the supplementary material. Similar to the uni-directional case \citep{guo2018confidence}, we can show that $\beta_{X\to Y}$ is identifiable under the plurality rule in the existence of reverse causal effects.

We next study the identification of $\beta_\YtoX$ given that $\beta_{\XtoY}$ has been identified.
Under model \eqref{eqn:sem}, we show in Section 
S1.2
of 
the supplementary material  that
the model assumptions of (\ref{eqn:sem}) imply  that
\begin{align}\label{eqn:LinearReg}
	\cov\big\{ (X_i- \beta_{Y\rightarrow X} Y_i)R_{Y,i},Z_i\}=0.
\end{align}
Hence, if $\cov(Y_iR_{Y,i} ,Z_i)\neq 0$, then  $\beta_{Y\rightarrow X}$ is identifiable based on (\ref{eqn:LinearReg}).
We term $\cov(Y_iR_{Y,i} ,Z_i)\neq 0$ as the  covariance heterogeneity (CH) condition for the direction $\YtoX$.
The CH condition essentially requires some heteroscedasticity in the error term $\zeta_i$ of model~\eqref{eqn:sem}. 
Heteroscedasticity has been widely utilized in previous studies for identification in linear structural models without exclusion restrictions \citep{lewbel2012using, tchetgen2021genius, sun2022selective}. Under our model assumptions, as shown in Section S1.3
of the supplementary material, we have
$
\cov( Y_iR_{Y,i}, Z_i) = (1-\beta_\XtoY\beta_\YtoX)^{-1}E(\zeta_i^2 Z_i)
$.
Consequently, if $E(\zeta_i^2 Z_i) \neq 0$, the nonzero covariance condition is satisfied. Below we provide an example in which the CH condition holds.
\begin{example}\label{exam:nonzerocov-prop}
	Suppose that the error term $\zeta_i$ in \eqref{eqn:sem} takes the form: $\zeta_i=\zeta_i^* Z_{i,1}$, where $\zeta_i^*\indep (\eta_i, U_i, Z_i)$, and $E(\zeta_i^*)=0$. If we additionally assume that $E(Z_{i,1}^3) \neq 0$, then the CH condition for $\YtoX$  is met.
\end{example}

The nonzero third moment of $Z_{i,1}$ in Example~\ref{exam:nonzerocov-prop} can be satisfied by many asymmetric distributions, such as chi-square, log-normal, and various discrete distributions. In particular, SNPs are often chosen as candidate IVs in MR studies, which typically take on three values. This scenario easily meets the nonzero third moment condition, as demonstrated in our simulation studies.
Below we state the identification results in bi-directional MR models under the aforementioned assumptions.
\begin{lemma}\label{lem:identification}
	Suppose that the plurality rule for  $\XtoY$ in \eqref{eqn:plurality} holds and $\cov(Y_iR_{Y,i} ,Z_i)\neq 0$,
	then $\beta_{X\rightarrow Y}$ and $\beta_{Y\rightarrow X}$ are identifiable.
\end{lemma}

Lemma~\ref{lem:identification} shows identification under the plurality rule  for $\XtoY$ and the CH condition for $\YtoX$. Similarly, if the plurality rule holds for $\YtoX$ and the CH condition for $\XtoY$ ($\cov(X_iR_{X,i},Z_i)\neq 0$)  holds,
then $\beta_{\XtoY}$ and $\beta_{\YtoX}$ are also identifiable.

In practical scenarios, however, it can be challenging to  determine the precise direction for which the  plurality rule holds without strong prior knowledge. 
A more realistic case is  only assuming  the plurality rule holds for one direction without specifying the detailed direction information.	
In this challenging scenario, the conditions in Lemma \ref{lem:identification} cannot guarantee the identification results.  
We provide a counter example in Section 
S1.4
of the supplementary material.
Therefore, it is necessary to enhance the plurality rule. We formally present the assumptions for bi-directional causal effect identification.

\begin{assumption}\label{ass:plurality}
	The enhanced plurality rule holds for $\XtoY$:
	\begin{align*}
		|\calV_\XtoY|>\max\bigg(\max_{c}\Big|\Big\{j\in\calV_{pl}\cap \calS_X:\frac{\pi_{Y,j}}{\gamma_{X,j}}=c \Big\} \Big|,|\calV_{\YtoX}|\mathbbm{1}(\beta_{\YtoX}\neq 0),|\calV_{pl}\setminus \calS_X|\mathbbm{1}(\beta_{\YtoX}\neq 0)\bigg).
	\end{align*}
	The covariance heterogeneity condition holds for $\YtoX$:
	$\cov(Y_iR_{Y,i},Z_i)\neq 0$.
\end{assumption}
Compared with the original plurality rule \eqref{eqn:pluralityrewrite}, the enhanced plurality rule additionally requires the set of valid IVs for $\XtoY$, $\calV_{\XtoY}$, to be larger another set $\calV_{pl}\setminus \calS_X$ when $\beta_{\YtoX}\neq 0$.
If $\beta_{\YtoX}=0$, i.e., there is no reverse causal effect, the enhanced plurality rule reduces to the conventional one in (\ref{eqn:pluralityrewrite}).
When $\beta_{\YtoX}\neq 0$, the enhanced plurality rule is stronger. Indeed, the variables in $\calV_{pl}\setminus \calS_X$ are irrelevant IVs for the direction $\XtoY$ but relevant invalid IVs for the direction $\YtoX$ with equal ratios $\pi_{X,j}/\gamma_{Y,j}=-\beta_{\YtoX}$. If this set of IVs is the largest cluster, they can be mis-identified as valid IVs for $\YtoX$ and one may mis-believe the plurality rule holds for the direction $\YtoX$. By restricting $\calV_{\XtoY}$ to be the largest cluster, we avoid this confusion at the population level.

In practice, we often have no prior knowledge about which direction the enhanced plurality rule holds for. Hence, we provide a parallel version of Assumption \ref{ass:plurality} by switching the order of $X$ and $Y$.

\begin{assumptionp}{\ref{ass:plurality}$'$}
	\label{ass:pluralityYtoX}
	The enhanced plurality rule holds for $\YtoX$:
	\begin{align*}
		|\calV_\YtoX|>\max\bigg(\max_{c}\Big|\Big\{j\in\calV_{pl}\cap \calS_Y:\frac{\pi_{X,j}}{\gamma_{Y,j}}=c \Big\} \Big|,|\calV_{\XtoY}|\mathbbm{1}(\beta_{\XtoY}\neq 0),|\calV_{pl}\setminus \calS_{Y}|\mathbbm{1}(\beta_{\XtoY}\neq 0)\bigg).
	\end{align*}
	The covariance heterogeneity condition holds for $\XtoY$:
	$\cov(X_iR_{X,i},Z_i)\neq 0$.
\end{assumptionp}

Below we discuss the identification of $\beta_\XtoY$ and $\beta_\YtoX$ for the bi-directional case when only one of Assumption \ref{ass:plurality} and Assumption \ref{ass:pluralityYtoX} is true, without knowing which one holds. 
Specifically, as in the one-directional case, the enhanced plurality rule motivates a mode-based method to identify one causal effect, say, $\beta_{\XtoY}$. Once $\beta_{X\rightarrow Y}$ is identified, it follows from equation~\eqref{eqn:LinearReg} that 
\begin{align*}
	X_i R_{Y,i}=\beta_{Y\rightarrow X} Y_i R_{Y,i}+\mu_i,~\cov(\mu_i, Z_i)=0.
\end{align*}
The CH condition for $\YtoX$ implies that $\text{Cov}(Y_i R_{Y,i},Z_i)\neq 0$, which motivates a two-stage least square (TSLS) method to estimate $\beta_{Y\rightarrow X}$.  The whole pipeline is termed PCH method, which is short for the ``plurality-then-covariance-heterogeneity''. We first introduce the oracle PCH method, which is defined based on the population moments of the observed data.

\begin{definition}[The oracle PCH method]\label{def:pch-estimand}
	Define $(\beta_{D\rightarrow D'}^\star,\beta_{D'\rightarrow D}^\star, \calV^{\star}_{D\rightarrow D'})=\textup{oPCH}(D,D')$ as follows. Given $\gamma_D$ and $\gamma_{D'}$, let
	\[
	b_{D\rightarrow D'}\in\argmax_{b\in\mathbbm{R}}\big|\big\{j\in \calS_D:\frac{\gamma_{D',j}}{\gamma_{D,j}}=b\big\}\big|.
	\]
	If $b_{D\rightarrow D'}\neq b'_{D\rightarrow D'}$ are both solutions to the above optimization, then define $\beta_{D\rightarrow D'}^\star=\infty$ and $\calV^{\star}_{D\rightarrow D'}=\emptyset$;
	if $b_{D\rightarrow D'}$ is the unique solution to the above optimization, then $\beta_{D\rightarrow D'}^\star= b_{D\rightarrow D'}={\rm mode} \{\gamma_{D',j}/\gamma_{D,j},j\in \calS_D\}$ and $\calV^{\star}_{D\rightarrow D'}=\{j\in \calS_D: \gamma_{D',j}/\gamma_{D,j}=\beta_{D\rightarrow D'}^\star\}$.	
	
	Let $\bar D'=D'-\beta_{D\rightarrow D'}^\star D$ and $\Lambda=\bar D'-E(\bar D'\mid Z)$. Define
	\begin{align*}
		\theta_D=\Sigma^{-1} E(\Lambda_{i}D_iZ_i),\quad \text{and}\quad \theta_{D'}=\Sigma^{-1} E(\Lambda_{i}D'_iZ_i).
	\end{align*}
	If $\theta_{D'}^\T\Sigma\theta_{D'}\neq 0$, define $\beta_{D'\rightarrow D}^\star=(\theta_{D}^\T\Sigma\theta_{D'})/(\theta_{D'}^\T\Sigma\theta_{D'})$;
	if $\theta_{D'}^\T\Sigma\theta_{D'}= 0$, define $\beta_{D'\rightarrow D}^\star=\infty$.
\end{definition}

Definition \ref{def:pch-estimand} is a population-level algorithm which applies the mode-based method in one direction and applies CH-based method in the other direction.
The mode-based method may be viewed as the population-level counterpart of TSHT. However, our definition involves an extra step to check whether the mode is unique. If the mode $b_{D\rightarrow D'}$ is not unique, then we know that the enhanced plurality rule for $D\rightarrow D'$ is  violated. For the CH method, we consider the population-level of TSLS based on (\ref{eqn:LinearReg}).
When applying the CH rule, we also check whether the denominator is zero or not. If $\theta_{D'}=0$, then the CH assumption cannot hold for the direction $D'\rightarrow D$. These two checking steps are crucial for achieving identification in the uni-directional case as we will see in Section \ref{subsec:uni-identification}.
In practice, we have no prior knowledge of which direction the plurality rule holds. Hence, we need to apply the oPCH method by switching the two traits.
We denote 
\begin{align}
	\label{eq:pch-out}
	(\beta_{\XtoY,\I}, \beta_{\YtoX,\I},\calV_{\XtoY,\I})=\textup{oPCH}(X,Y)~\text{and}~(\beta_{\YtoX,\II}, \beta_{\XtoY,\II},\calV_{\YtoX,\II})=\textup{oPCH}(Y,X).
\end{align}
We see that these two triplets are population-level oracles defined based on $\gamma_X,\gamma_{Y}$, and the moments of observed variables. 
In the next theorem, we show that these two triplets identify the true causal effects to certain extent.

\begin{theorem}\label{thm:bi-ident}
	Suppose that  either Assumption~\ref{ass:plurality} or Assumption~\ref{ass:pluralityYtoX} holds. 
	If $\beta_{\XtoY}\beta_{\YtoX}\neq 0$, then 
	$(\beta_\XtoY,\beta_\YtoX)$ is identifiable up to two points $(\beta_{\XtoY,\I},\beta_{\YtoX,\I})$ and $(\beta_{\XtoY,\II},\beta_{\YtoX,\II})$. Moreover, 
	\begin{align*}
		&\beta_{\XtoY,\I}=1/\beta_{\YtoX,\II},~~\beta_{\YtoX,\I}=1/\beta_{\XtoY,\II}.
	\end{align*}
\end{theorem}
Theorem \ref{thm:bi-ident} implies that one of $(\beta_{\XtoY,\I},\beta_{\YtoX,\I})$ and $(\beta_{\XtoY,\II},\beta_{\YtoX,\II})$ equals  $(\beta_\XtoY,\beta_\YtoX)$ and the other pair equals $(1/\beta_\YtoX,1/\beta_\XtoY)$.
This conclusion implies that
these two pairs are indistinguishable if  one of Assumptions \ref{ass:plurality} and \ref{ass:pluralityYtoX} holds and $\beta_{\XtoY}\beta_{\YtoX}\neq 0$.
The main reason is that the enhanced plurality rule for two directions cannot hold simultaneously, but we do not know which one holds. 
When some prior knowledge about $(\beta_\XtoY,\beta_\YtoX)$ is further available, we can uniquely identify the causal parameters. For instance, in the case where $\beta_\XtoY$ and $\beta_\YtoX$ have different signs, if the information about their signs are known by researchers, which may be available in many studies, then $\beta_\XtoY$ and $\beta_\YtoX$ are uniquely identified, as shown in the following corollary.

\begin{corollary}
	\label{cor:bi-unique}
	Suppose that  either Assumption~\ref{ass:plurality} or Assumption~\ref{ass:pluralityYtoX} holds. 
	If $\beta_{\XtoY}>0$ and $\beta_{\YtoX}<0$, then we have
	\[
	\beta_\XtoY=\max(\beta_{\XtoY,\I},\beta_{\XtoY,\II})~\text{and}~\beta_{\YtoX}=\min(\beta_{\YtoX,\I},\beta_{\YtoX,\II}).
	\]
\end{corollary}
In the case where $\beta_\XtoY<0$ and $\beta_\YtoX>0$, similar identification results can be obtained as in Corollary~\ref{cor:bi-unique}. Besides, other prior knowledge can also be used to uniquely determine $\beta_\XtoY$ and $\beta_\YtoX$. For instance, if we know that the relationships between $X$ and $Y$ are not too strong, such as $0<|\beta_\XtoY|<1,0<|\beta_\YtoX|<1$, then $\beta_\XtoY$ and $\beta_\YtoX$ are also identifiable.

\subsection{Uni-directional causal identification}
\label{subsec:uni-identification}
Next, we investigate the identification of causal effects when $\beta_{\XtoY}\beta_{\YtoX}=0$.
We will show that the causal effects can be identified based on the two triplets in \eqref{eq:pch-out} when $\beta_{\XtoY}\beta_{\YtoX}=0$.
\begin{theorem}
	\label{thm:uni-ident}
	Suppose that  either Assumption~\ref{ass:plurality} or Assumption~\ref{ass:pluralityYtoX} holds. 
	If $\beta_{\XtoY}\beta_{\YtoX}=0$, then $(\beta_{\XtoY},\beta_{\YtoX})$ is identifiable.
\end{theorem}
Let $\bm{b}=(\beta_{\XtoY,\I},\beta_{\YtoX,\I},\beta_{\XtoY,\II},\beta_{\YtoX,\II})^\T$. 
In the proof of Theorem \ref{thm:uni-ident}, we show that under its conditions, $(\beta_{\XtoY},\beta_{\YtoX})=(\beta^{\circ}_{\XtoY},\beta^{\circ}_{\YtoX})$ for
\begin{align}
	\label{pch-ident2}
	(\beta^{\circ}_{\XtoY},\beta^{\circ}_{\YtoX})=\left\{\begin{array}{ll}
		( \beta_{\XtoY,\I},\beta_{\YtoX,\I}  )   & \max(\beta_{\XtoY,\II},\beta_{\YtoX,\II})=\infty,\\
		( \beta_{\XtoY,\II},\beta_{\YtoX,\II}  )   &  \max(\beta_{\XtoY,\I},\beta_{\YtoX,\I})=\infty,\\
		( \beta_{\XtoY,\I},\beta_{\YtoX,\I}  )   &  \text{if}~|\calV_{\XtoY,\I}|\geq |\calV_{\YtoX,\II}|~\text{and}~\max(\bm{b})<\infty,\\
		( \beta_{\XtoY,\II},\beta_{\YtoX,\II}  )   &   \text{if}~|\calV_{\XtoY,\I}|<|\calV_{\YtoX,\II}|~\text{and}~\max(\bm{b})<\infty.
	\end{array}\right.
\end{align}
We see that $(\beta^{\circ}_{\XtoY},\beta^{\circ}_{\YtoX})$ are defined solely based on the estimands computed via \eqref{eq:pch-out}. As we mentioned before, these estimands are functions of $\gamma_X$, $\gamma_Y$, and the moments of observed variables. Hence, the fact that $(\beta_{\XtoY},\beta_{\YtoX})=(\beta^{\circ}_{\XtoY},\beta^{\circ}_{\YtoX})$ implies the identifiability of the causal effects.
The idea of the identification rule in (\ref{pch-ident2}) is as follows. 
If $\max(\beta_{\XtoY,\II},\beta_{\YtoX,\II})=\infty$, then by Definition \ref{def:pch-estimand}, we know that either the enhanced plurality rule for $\YtoX$ does not hold or the CH condition for $\XtoY$ does not hold, which implies that Assumption \ref{ass:pluralityYtoX} is violated. Hence, Assumption \ref{ass:plurality} must be true and $ ( \beta_{\XtoY,\I},\beta_{\YtoX,\I})$ identifies the true causal effects. We see that the two checking steps provide insights in determining whether Assumption \ref{ass:plurality} or Assumption \ref{ass:pluralityYtoX} is true.
If $\max(\bm{b})<\infty$, we compare $|\calV_{\XtoY,\I}|$ and $|\calV_{\YtoX,\II}|$, which can also tell us which one of Assumptions \ref{ass:plurality} and \ref{ass:pluralityYtoX} holds thanks to the enhanced plurality rule. 

Theorem \ref{thm:uni-ident} implies that the estimands of oralce PCH can be used to identify the causal effects in the uni-directional case even if the causal direction is unknown a priori. Together with Theorem \ref{thm:bi-ident}, we have studied identification of causal effects in the bi-directional and uni-directional cases separately. As we do not know whether the true causal relationship is bi-directional or not, we study the identification of the causal direction in the next section.

\subsection{Causal direction identification}
\label{subsec:direction-ident}
In this subsection, we study the identification of the causal direction. Denote the true causal direction by
\begin{align*}
	\mathcal{H}=\left\{
	\begin{array}{ll}
		2 ~&\text{if}~\beta_{\XtoY}\beta_{\YtoX}\neq 0,\\
		1 ~&\text{if} ~\beta_{\XtoY}\neq 0,\beta_{\YtoX}=0,\\
		0~&\text{if}~\beta_{\XtoY}=\beta_{\YtoX}=0,\\
		-1 ~&\text{if}~\beta_{\XtoY}= 0,\beta_{\YtoX}\neq 0.
	\end{array}\right.
\end{align*}
We see that $\mathcal{H}$ encompasses all the possible causal relationships between two variables corresponding to Figure \ref{fig:dags-model}. In the next theorem, we show that $\mathcal{H}$ can be identified based on the two triplets  (\ref{eq:pch-out}) defined by oracle PCH.
Define
\begin{align*}
	\mathcal{H}^{\circ}=\left\{
	\begin{array}{ll}
		2 ~&\text{if}~\min_{j\leq 4}|\bm{b}_j|>0,\\
		1 ~&\text{if} ~\beta^{\circ}_{\XtoY}\neq 0,~\beta^{\circ}_{\YtoX}=0,\\
		0~&\text{if}~\beta^{\circ}_{\XtoY}=\beta^{\circ}_{\YtoX}=0,\\
		-1 ~&\text{if}~\beta^{\circ}_{\XtoY}=0,~\beta^{\circ}_{\YtoX}\neq 0.
	\end{array}\right.
\end{align*}
Note that $\mathcal{H}^{\circ}$ is defined based on $\bm{b}$ and $(\beta_{\XtoY}^{\circ},\beta_{\YtoX}^{\circ})$, which are functions of the output of oracle PCH.
Hence, $\mathcal{H}^{\circ}$ is also a population-level statistic of the data.
\begin{theorem}
	\label{cor:direction-ident}
	Assume that Assumption \ref{ass:plurality} or Assumption \ref{ass:pluralityYtoX} holds. Then the causal direction $\mathcal{H}$ is identifiable and $\mathcal{H}=\mathcal{H}^{\circ}$.
\end{theorem}
Theorem \ref{cor:direction-ident} provides an identification method for the causal direction. We first mention that the four cases considered in $\mathcal{H}^{\circ}$ provide a mutually exclusive and exhaustive partition of the whole space. To see this, if $\min_{j\leq 4}|\bm{b}_j|=0$, Theorem \ref{thm:bi-ident} implies that it must hold that $\beta_{\XtoY}\beta_{\YtoX}=0$. As we show in the proof of Theorem \ref{thm:uni-ident}, $(\beta_{\XtoY}^{\circ},\beta_{\YtoX}^{\circ})$ identifies the true causal effect when $\beta_{\XtoY}\beta_{\YtoX}=0$. This explains why we determine the causal direction based on $\beta^{\circ}_{\XtoY}$ and $\beta^{\circ}_{\YtoX}$ when $\min_{j\leq 4}|\bm{b}_j|=0$.
It is also worth noticing that even if the causal effects cannot be uniquely identified in the bi-directional case, we can identify $\calH$ without extra conditions. This is because the two pairs of PCH estimands are both nonzero when $\beta_{\XtoY}\beta_{\YtoX}\neq 0$ under the conditions of Theorem \ref{thm:bi-ident}.

Theorems \ref{thm:bi-ident}--\ref{cor:direction-ident} together provide a complete identification result of the casual effects and the casual directions.
If $\mathcal{H}=2$, then one can use Theorem \ref{thm:bi-ident} and Corollary \ref{cor:bi-unique} to identify the causal effects. If $\mathcal{H}<2$, then one can identify the true casual effects and causal directions based on $(\beta^{\circ}_{\XtoY},\beta^{\circ}_{\YtoX})$.

In the next section, we introduce  a finite-sample method for causal inference and causal discovery based on these identification results.

\section{Proposed method}
\label{sec:method}

In this section, we introduce empirical methods for estimating the causal directions and causal effects. In Section \ref{subsec:fs-pch}, we introduce the finite-sample PCH method.  In Section \ref{subsec:inf}, we present the final algorithm to make inference of the causal directions and effects. In Section \ref{sec:theory}, we provide theoretical guarantees for the proposed method.

\subsection{Finite-sample PCH method}
\label{subsec:fs-pch}
We first introduce the empirical version of the oracle PCH given in Definition \ref{def:pch-estimand}, shorthanded as PCH. The PCH method also has two main steps.
The first step estimates one causal effect based on the enhanced plurality rule. For this purpose, we consider a modified TSHT method with a diagnostic step checking whether the mode is unique. The second step estimates the other causal effect with CH method, where we also add a diagnostic step to check whether the CH condition holds.
The PCH method is summarized in Algorithm \ref{alg:pch}.

\begin{algorithm}
	{\bf Input:}   $D$, $D'$, $Z$.
	
	{\bf  Output:} $\{(\hat{\beta}_{D\to D'},\hat{\sigma}^2_{D\to D'},\hat{\beta}_{D'\to D},\hat{\sigma}^2_{D'\to D}),\hat{\calV}_{D\to D'}\}$.
	
	Compute $\hat{\gamma}_D=(Z^\T Z)^{-1}Z^\T D$ and $\hat{\gamma}_{D'}=(Z^\T Z)^{-1}Z^\T D'$.

	Compute $\hat{\calS}_D=\{j: |\hat{\gamma}_{D,j}|\geq \hat{\sigma}_{D,j}\sqrt{\log n/n}\}$, where $\hat{\sigma}_{D,j}$ is defined in (S21) of the supplementary material. Compute the votes received by each IV in $\hat{\calS}_D$ as follows
	\[
	\hat{Q}_j=\left|\left\{k\in\hat{\calS}_D: \bigg|\frac{\hat{\gamma}_{D',k}}{\hat{\gamma}_{D,k}}-\frac{\hat{\gamma}_{D',j}}{\hat{\gamma}_{D,j}}\bigg|\leq \hat{\sig}_{k\rightarrow j}\sqrt{\frac{\log n}{n}}\right\}\right|,
	\]
	where $\hat{\sig}_{k\rightarrow j}$ is defined in (S22) of the supplementary material.
	Let 
	\[
	\tilde{\calV}_{D\to D'}=\left\{j\in \hat{\calS}_D:\hat{Q}_j=\max_{k\in\hat{\calS}_D}\hat{Q}_k\right\}.
	\]
	
	\eIf{ $|\tilde{\calV}_{D\to D'}|= \max_{j\in\hat{\calS}_D}\hat{Q}_j$}{ Set $\hat{\calV}_{D\to D'}=\tilde{\calV}_{D\to D'}$. Obtain $(\hat{\beta}_{D\to D'},\hat{\sigma}^2_{D\to D'})$ by performing modified TSLS with exposure $D$, outcome $D'$, IVs $Z_{\cdot,\hat{\calV}_{D\to D'}}$ and measured confounders $Z_{\cdot,\hat{\calV}^c_{D\to D'}}$ as in 
		(S23) and (S24)
		of the supplementary material.
		
		Let $\bar{D}'=D'-\hat{\beta}_{D\to D'}D$ and
		$\hat \Lambda = P_Z^{\perp} \bar{D}'$. Compute
		$ \hat\theta_D=(Z^\T Z)^{-1}Z^\T (D\odot \hat \Lambda)$ and $\hat\theta_{D'}=(Z^\T Z)^{-1}Z^\T (D'\odot \hat \Lambda)$.

		\eIf{ $\hat{\theta}_{D'}^\T\hat{\Omega}_{D'}\hat{\theta}_{D'}\geq \chi^2_p(1-1/n)$ for $\hat{\Omega}_{D'}$ defined in 
			(S26)
		}{
			Estimate the causal effect $\beta_{D'\to D}$ by 
			$\hat\beta_{D'\to D}=({\hat\theta}_{D'}^\T \hat\Sigma \hat\theta_{D})/({\hat\theta}_{D'}^\T \hat\Sigma \hat\theta_{D'}).$
			The variance of $\hat{\beta}_{D'\to D}$, $\hat{\sigma}^2_{D'\to D}$, can be estimated via 
			(S27).
		}{
			Set $\hat{\beta}_{D'\to D}=\hat{\sigma}^2_{D'\to D}=\textup{NA}$.}

	}{ Set $\hat{\calV}_{D\to D'}=\emptyset$, $\hat{\beta}_{D\to D'}= \hat{\sigma}^2_{D\to D'} =\hat{\beta}_{D'\to D}= \hat{\sigma}^2_{D'\to D} =\text{NA}$.}
	
	\caption{PCH method (PCH($D,D',Z$)).}
	\label{alg:pch}
\end{algorithm}

Line 5 of the PCH method includes a step detecting whether the mode is unique, which can signal the failure of the enhanced plurality rule for $D\rightarrow D'$. This step is not considered in the original TSHT as the causal relationship is assumed to be one-directional with known direction in their setting. In contrast, under Assumption \ref{ass:plurality} or Assumption \ref{ass:pluralityYtoX}, we need to detect whether the plurality rule holds.
If the mode is unique, the causal estimate is valid and we further provide it uncertainty quantification. The variance estimate  of $\hat{\beta}_{D\rightarrow D'}$ in 
(S24)
of the supplementary material is computed based on the empirical noises rather than the limiting distributions. This relaxation is important in the current problem because we consider the existence of covariance heterogeneity, i.e., $R_{Y,i}$ and $R_{X,i}$ are heteroscedastic given $Z_i$. 

The CH estimators are realized in line 7 to line 12.
We first test $H_0: \theta_{D'}=0$ in line 8, where $\chi^2_d(\alpha)$ is the $\alpha$ quantile of the $\chi^2$ distribution with degrees of freedom $d$. This step can signal the failure of the CH condition for $D'\to D$ if $\hat{\theta}_{D'}^\T\hat{\Omega}_{D'}\hat{\theta}_{D'}$ is not significantly larger than zero. 
If the CH condition is satisfied, we obtain the estimate $\hat{\beta}_{D' \to D}$ based on the two-stage least square with exposure $D'\odot \hat \Lambda$ and outcome $D\odot \hat \Lambda$. Its variance estimate $\hat{\sig}^2_{D'\to D}$ in 
(S27)
of the supplementary material is also based on the empirical noises.

We see that finite-sample versions of $(\beta_{\XtoY,\I},\beta_{\YtoX,\I},\calV_{\XtoY,\I})$ and $(\beta_{\YtoX,\II},\beta_{\XtoY,\II},\calV_{\YtoX,\II})$ can be realized based on Algorithm \ref{alg:pch}. 
Define 
\begin{align*}
	&\{(\hat{\beta}_{\XtoY,\I},\hat{\sigma}^2_{\XtoY,\I},\hat{\beta}_{\YtoX,\I},\hat{\sigma}^2_{\YtoX,\I}), \hat{\calV}_{\XtoY,\I}\}=\textup{PCH}(X,Y,Z)~\text{and}\\
	&\{(\hat{\beta}_{\YtoX,\II},\hat{\sigma}^2_{\YtoX,\II},\hat{\beta}_{\XtoY,\II},\hat{\sigma}^2_{\XtoY,\II}) ,\hat{\calV}_{\YtoX,\II}\}=\textup{PCH}(Y,X,Z).
\end{align*}
In the next subsection, we finish the causal inference of the causal directions and effects based on the above two tuples of estimates.

\subsection{Causal inference method}
\label{subsec:inf}

In this subsection, we finalize the proposed method for making bi-directional causal inference based on the output of Algorithm \ref{alg:pch}. The outline is as follows. First, it determines whether the causal relationship is bi-directional based on Theorem \ref{cor:direction-ident}. If the estimated causal relationship is bi-directional, it applies Corollary \ref{cor:bi-unique} to make bi-directional causal inference. If the estimated causal relationship is uni-directional, then it applies Theorem \ref{thm:uni-ident} to make inference of the causal effects. The formal procedure is summarized in Algorithm \ref{alg:inf}. 

\begin{algorithm}
	\textbf{Input:} $\{(\hat{\beta}_{X\rightarrow Y,\I}, \hat{\sigma}^2_{X\rightarrow Y,\I}, \hat{\beta}_{Y\rightarrow X,\I}, \hat{\sigma}^2_{Y\rightarrow X,\I}),\hat{\calV}_{\XtoY,\I}\}$,
	$\{(\hat{\beta}_{\YtoX,\II}, \hat{\sigma}^2_{\YtoX,\II}, \hat{\beta}_{\XtoY,\II}, \hat{\sigma}^2_{\XtoY,\II}),\hat{\calV}_{\YtoX,\II}\}$.
	
	\textbf{Output:} $\hat\beta_\XtoY$, $\hat\beta_\YtoX$, $\widehat{\CI}_{\XtoY}(\alpha)$, 
	$\widehat{\CI}_{\YtoX}(\alpha)$, $\widehat{\calH}$.

	
	\For{``$*$'' $\in\{``\XtoY,\I",``\YtoX,\I", ``\XtoY,\II",``\YtoX,\II"\}$}{ Compute
		$\widehat{\CI}_{*}(\alpha)=\big(\hat\beta_{*}-z_{1-\alpha/2}\hat\sigma_{*}/\sqrt{n},~ \hat\beta_{*}+z_{1-\alpha/2}\hat\sigma_{*}/\sqrt{n}\big).$ 
	}
	
	Compute $\overline\CI(\alpha)=\widehat{\CI}_{\XtoY,\I}(\alpha)\cup \widehat{\CI}_{\YtoX,\I} (\alpha)\cup\widehat{\CI}_{\XtoY,\II}(\alpha)\cup \widehat{\CI}_{\YtoX,\II}(\alpha)$. 
	
	\eIf{$\overline{\CI}(1/n)=\textup{NA}$~\text{or}~$0\in\overline{\CI}(1/n)$}{
		\uIf{$\textup{NA}\in\{\hat{\beta}_{\XtoY,\II},\hat{\beta}_{\YtoX,\II}\}$}{
			$\hat\beta_\XtoY=\hat\beta_{\XtoY,\I}$, $\widehat{\CI}_{\XtoY}(\alpha)=\widehat{\CI}_{\XtoY,\I}(\alpha)$, \\ $\hat\beta_\YtoX=\hat\beta_{\YtoX,\I}$, $\widehat{\CI}_{\YtoX}(\alpha)=\widehat{\CI}_{\YtoX,\I}(\alpha)$.
		}
		\uElseIf{
			$\textup{NA}\in\{\hat{\beta}_{\XtoY,\I},\hat{\beta}_{\YtoX,\I}\}$
		}{
			$\hat\beta_\XtoY=\hat\beta_{\XtoY,\II}$, $\widehat{\CI}_{\XtoY}(\alpha)=\widehat{\CI}_{\XtoY,\II}(\alpha)$, \\ $\hat\beta_\YtoX=\hat\beta_{\YtoX,\II}$, $\widehat{\CI}_{\YtoX}(\alpha)=\widehat{\CI}_{\YtoX,\II}(\alpha)$.
		}
		\uElseIf{$|\hat{\calV}_{\XtoY,\I}|\geq |\hat{\calV}_{\YtoX,\II}|$}{
			$\hat\beta_\XtoY=\hat\beta_{\XtoY,\I}$, $\widehat{\CI}_{\XtoY}(\alpha)=\widehat{\CI}_{\XtoY,\I}(\alpha)$, \\ $\hat\beta_\YtoX=\hat\beta_{\YtoX,\I}$, $\widehat{\CI}_{\YtoX}(\alpha)=\widehat{\CI}_{\YtoX,\I}(\alpha)$.
		}
		
		\Else{
			$\hat\beta_\XtoY=\hat\beta_{\XtoY,\II}$, $\widehat{\CI}_{\XtoY}(\alpha)=\widehat{\CI}_{\XtoY,\II}(\alpha)$, \\ $\hat\beta_\YtoX=\hat\beta_{\YtoX,\II}$, $\widehat{\CI}_{\YtoX}(\alpha)=\widehat{\CI}_{\YtoX,\II}(\alpha)$.
		}
		$\widehat{\calH}=\mathbbm{1}(0\not\in\widehat{\CI}_{\XtoY}(1/n))-\mathbbm{1}(0\not\in\widehat{\CI}_{\YtoX}(1/n))$.
	}{
		$\widehat{\calH}=2$.
		
		\eIf{$\hat{\beta}_{\XtoY,\I}>\hat{\beta}_{\XtoY,\II}$}{
			$\hat\beta_\XtoY=\hat\beta_{\XtoY,\I}$, $\widehat{\CI}_{\XtoY}(\alpha)=\widehat{\CI}_{\XtoY,\I}(\alpha)$, \\ $\hat\beta_\YtoX=\hat\beta_{\YtoX,\I}$, $\widehat{\CI}_{\YtoX}(\alpha)=\widehat{\CI}_{\YtoX,\I}(\alpha)$.
		}{
			$\hat\beta_\XtoY=\hat\beta_{\XtoY,\II}$, $\widehat{\CI}_{\XtoY}(\alpha)=\widehat{\CI}_{\XtoY,\II}(\alpha)$, \\ $\hat\beta_\YtoX=\hat\beta_{\YtoX,\II}$, $\widehat{\CI}_{\YtoX}(\alpha)=\widehat{\CI}_{\YtoX,\II}(\alpha)$.
		}
	}
	\caption{Causal inference in bi-directional models.}
	\label{alg:inf}
\end{algorithm}

As shown in Algorithm \ref{alg:pch}, the output estimators can be $\textup{NA}$.
In  Algorithm \ref{alg:inf}, we use the convention that $\widehat{\CI}_*(\alpha)=\textup{NA}$ if its associated estimators involve NA and $\overline{\CI}(\alpha)=\textup{NA}$ if any of $\widehat{\CI}_*(\alpha)=\textup{NA}$.
Algorithm \ref{alg:inf} achieves inference of the causal effects and causal discovery simultaneously.
In line 7, we determine whether the causal relationship is bi-directional or not. Theorem \ref{cor:direction-ident} implies that we should consider $\min_{j\leq 4}|\bm{b}_j|>0$ or not at population level. Empirically, we use the finite-sample estimate of $\bm{b}$ and take its uncertainty into account. Hence, in line 7, we consider the union of four confidence intervals with confidence level $1-1/n$. The confidence level goes to one as $n\rightarrow\infty$, which can lead to consistent detection. If the causal relationship is determined to be uni-directional, lines 8 to 20 find the confidence intervals based on (\ref{pch-ident2}). If the causal relationship is determined to be bi-directional, we leverage the prior knowledge that $\beta_{\XtoY}>0$ in line 24 to uniquely estimate the causal effects.
The estimation of $\calH$ follows the same idea as in Theorem \ref{cor:direction-ident}.
We will prove the consistency of $\widehat{\calH}$ and provide corresponding theoretical results on inference in the next subsection.



\subsection{Theoretical guarantees}
\label{sec:theory}

In this subsection, we provide theoretical guarantees for the proposed method introduced in Section \ref{subsec:inf} including both the causal discovery and inference results.

\begin{theorem}[Consistent discovery]\label{thm:alg-consistency}
	Suppose that either Assumption~\ref{ass:plurality} or \ref{ass:pluralityYtoX} holds. Then 
	\begin{align*}
		\lim_{n\rightarrow \infty}\pr(\widehat\calH =\calH)=1.
	\end{align*}
\end{theorem}
Theorem~\ref{thm:alg-consistency} shows that the proposed method in Algorithm \ref{alg:inf} can consistently recover the causal direction between $X_i$ and $Y_i$. As introduced in Algorithm \ref{alg:inf}, our discovery method is based on the asymptotic properties of the estimators $\hat\beta_{\XtoY,\I}$, $\hat\beta_{\YtoX,\I}$, $\hat\beta_{\XtoY,\II}$, $\hat\beta_{\YtoX,\II}$. Details about the asymptotic results of these estimators under Assumption~\ref{ass:plurality} or \ref{ass:pluralityYtoX} are
provided in Section 
S3.2
of the supplementary material.

\begin{theorem}[Inference]\label{thm:inference}
	Suppose that either Assumption~\ref{ass:plurality} or \ref{ass:pluralityYtoX} holds. We further assume that $\beta_\XtoY>0$ and $\beta_\XtoY<0$ if $\beta_\XtoY\beta_\YtoX\neq 0$. Then
	\begin{align*}
		&\lim_{n\rightarrow\infty}\pr\{\beta_{\XtoY}\in \widehat\CI_{\XtoY}(\alpha)\}=1-\alpha,
		~~\text{and}~~
		\lim_{n\rightarrow\infty}\pr\{\beta_{\YtoX}\in \widehat\CI_{\YtoX}(\alpha)\}=1-\alpha.
	\end{align*}
\end{theorem}

Theorem~\ref{thm:inference} shows that the confidence intervals $\widehat\CI_{\XtoY}(\alpha)$ and $\widehat\CI_{\YtoX}(\alpha)$ output by 
Algorithm \ref{alg:inf} asymptotically cover the true causal effects $\beta_\XtoY$ and $\beta_\YtoX$ with level $1-\alpha$, respectively. Similar to Theorem~\ref{thm:alg-consistency}, this result also depends on the asymptotic properties of the estimators $\hat\beta_{\XtoY,\I}$, $\hat\beta_{\YtoX,\I}$, $\hat\beta_{\XtoY,\II}$, $\hat\beta_{\YtoX,\II}$. Indeed, we show that these estimators are asymptotically normal in Section 
S3.2
of the supplementary material. However, because $(\beta_\XtoY,\beta_\YtoX)$ is not uniquely identifiable under Assumption \ref{ass:plurality} or \ref{ass:pluralityYtoX} when $\beta_\XtoY\beta_\YtoX\neq 0$, additional conditions as given in Corollary~\ref{cor:bi-unique} are needed for inference in this case. As discussed below Corollary~\ref{cor:bi-unique}, such conditions can be replaced by other prior knowledge, as long as they can guarantee the unique identifiability of the two parameters, and similar inference results can also be established. 


Theorems~\ref{thm:alg-consistency} and \ref{thm:inference} establish the consistency of causal discovery and the validity of inference for the proposed method. In contrast, many existing causal discovery methods lack rigorous theoretical guarantees, particularly regarding the inference of the discovered relationships.
As far as we are aware, this
is the first paper to incorporate invalid IVs for simultaneous consistent discovery and valid inference on possibly bi-directional relationships between two traits, with the exception of \cite{chen2023discovery}.  They focus on the discovery of directed acyclic graph with unmeasured confounding, which is also applicable for cases involving only two traits.
However, their method is distinct from ours, because their identification and inference strategies rely on Gaussian model assumptions. They also do not consider
the bi-directional issue as explored here.

\section{Simulation}\label{sec:simulation}

In this section, we conduct simulation studies to evaluate the finite-sample performance of the proposed method. 
We compare our method with 
existing methods by separately applying TSHT, Egger, and inverse-variance weighted (IVW) estimation procedures to each direction. These three methods were originally proposed to estimate causal effects for the uni-directional relationship with known directions. Here we also use them to infer causal direction by checking whether the estimated confidence interval for each direction includes zero. For a fair comparison, we set the confidence level $1-\alpha$ to be the same as ours, i.e., $\alpha=1/n$, when inferring causal direction.

\begin{figure}[h!]
	\subfigure[$\beta_\XtoY=0$]{\includegraphics[width=\textwidth]{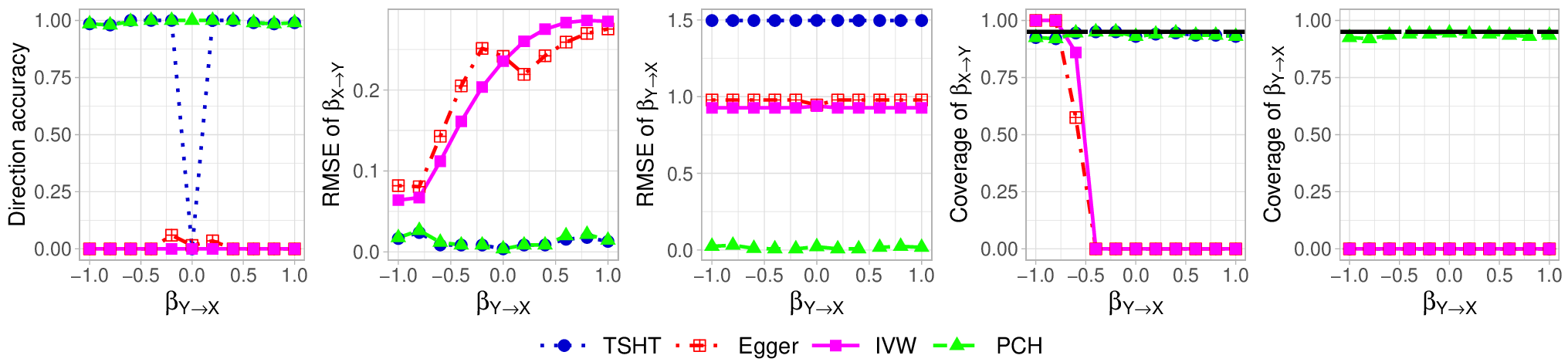}}
	\subfigure[$\beta_\YtoX=0$]{\includegraphics[width=\textwidth]{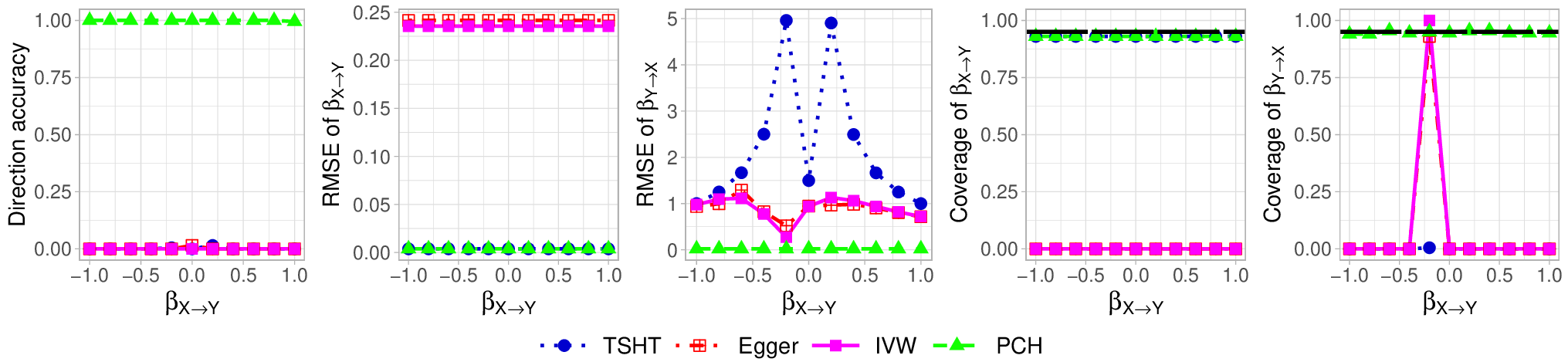}}
	\subfigure[$\beta_\XtoY=0.5$]{\includegraphics[width=\textwidth]{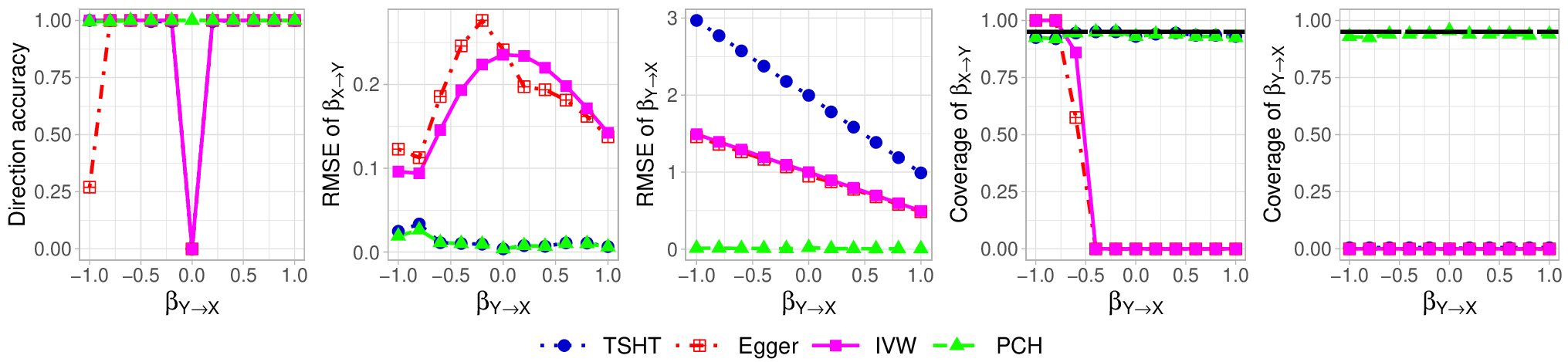}}
	\subfigure[$\beta_\YtoX=0.5$]{\includegraphics[width=\textwidth]{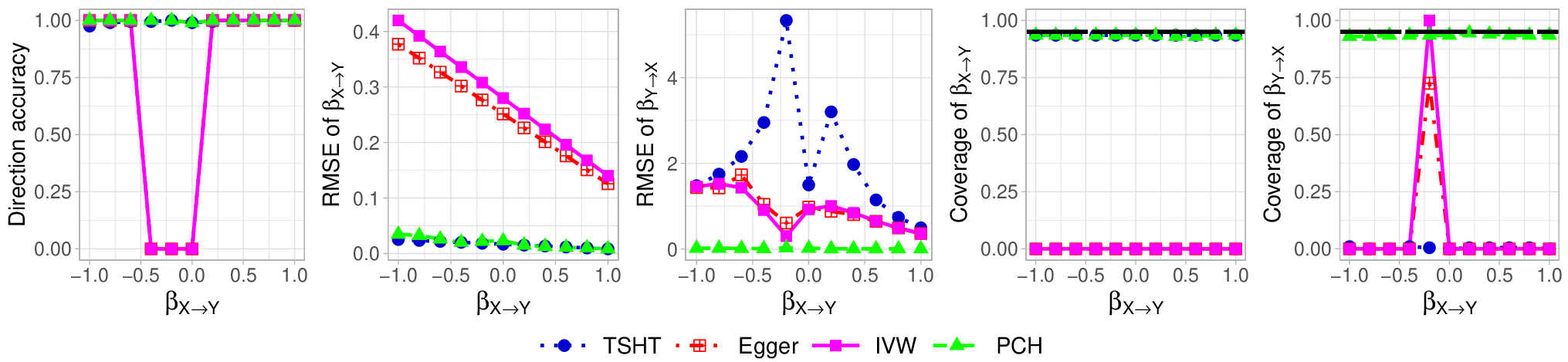}}
	\caption{Comparisons in cases (a)-(d)  between the proposed  (\texttt{PCH}) and  existing methods (\texttt{TSHT}, \texttt{Egger}, \texttt{IVW}) in terms of  direction detection accuracy, root mean squared error (RMSE) and 95\% coverage probability of estimating $\beta_\XtoY$ and $\beta_\YtoX$. The horizontal line in the right two plots marks the value 95\%.
	}\label{fig:sim-res}
\end{figure}

We first generate a vector of $p$ candidate instruments $Z_i\in\R^p$ with elements independently drawn from a three-level (1, 2, 3) discrete variable with probabilities (0.6, 0.2, 0.2). We then generate an unobserved confounding variable $U_i$ from a standard normal distribution $N(0,1)$. Finally, we generate the primary variables $Y_i$ and $X_i$ according to the following models:
\begin{align*}
	Y_i=& \beta_{X\rightarrow Y}X_i+\pi_Y^\T Z_i+U_i+\zeta_{i}, \\
	X_i=&\beta_{Y\rightarrow X}Y_i+\pi_X^\T Z_i+U_i+\eta_{i}.
\end{align*}
We set $\pi_Y,\pi_X$ as follows:
\begin{align*}
	\pi_Y=(\underbrace{0,\ldots,0}_{s_x},\underbrace{0.4,\ldots,0.4}_{s_{xy}+s_y},0,\ldots, 0)^\T,\quad\text{and}\quad \pi_X=(\underbrace{0.6,\ldots,0.6}_{s_x+s_{xy}},0,\ldots, 0)^\T,
\end{align*}
where $s_x=15$, $s_{xy}=8$, and $s_y=5$.
The error terms $\zeta_{i}$ and $\eta_i$ are generated as follows:
$\zeta_{i}=\zeta_{i}^*(Z_{i,s_x+1}+\tau Z_{i,s_x+2})$ and $\eta_i=\eta_{i}^*(Z_{i,1}+\kappa Z_{i,2})$, where $\zeta_{i}^*,\eta_{i}^*\sim N(0,1)$, $\tau=1-1.5\mathbbm{1}(\beta_\YtoX=0)$, and $\kappa=-1+2\mathbbm{1}(\beta_\YtoX=0)$.
Below we consider four cases of $(\beta_\XtoY,\beta_\YtoX)$.
\begin{align*}
	&\text{Case (a):}~~\beta_\XtoY=0,~~~~\beta_\YtoX\in\{-1,-0.8,\ldots,0.8,1\};\\
	&\text{Case (b):}~~\beta_\YtoX=0,~~~~\beta_\XtoY\in\{-1,-0.8,\ldots,0.8,1\};\\
	&\text{Case (c):}~~\beta_\XtoY=0.5,~~\beta_\YtoX\in\{-1,-0.8,\ldots,0.8,1\};\\
	&\text{Case (d):}~~\beta_\YtoX=0.5,~~\beta_\XtoY\in\{-1,-0.8,\ldots,0.8,1\}.
\end{align*}
In case (a), $\mathcal{H}=-1$ if $\beta_{\YtoX}\neq 0$ and $\mathcal{H}=0$ otherwise. In case (b), $\mathcal{H}=1$ if $\beta_{\XtoY}\neq 0$ and $\mathcal{H}=0$ otherwise. In case (c), $\mathcal{H}=2$ if $\beta_{\YtoX}\neq 0$ and $\mathcal{H}=1$ otherwise. In case (d), $\mathcal{H}=2$ if $\beta_{\XtoY}\neq 0$ and $\mathcal{H}=-1$ otherwise.
We choose the sample size  $n=50000$ and the dimension $p=100$. For each case, we report the accuracy of causal direction detection,  the root mean squared error (RMSE), and 95\% coverage probability for estimating $\beta_\XtoY$ and $\beta_\YtoX$. 
Figure~\ref{fig:sim-res} shows the simulation results for all methods averaged across 200 experiments in cases (a)-(d).

Figure \ref{fig:sim-res} shows that the proposed PCH method   outperforms all existing methods. The PCH method consistently estimates the causal direction between $X_i$ and $Y_i$ in all cases. Aditionally, it exhibits excellent performance in estimating both parameters $\beta_\XtoY$ and $\beta_\YtoX$, with the smallest RMSEs and reasonable coverage probabilities. In contrast, the performances of the other three naive methods, namely TSHT, Egger, and IVW, are unsatisfactory. For example, in case (a) of $\beta_\XtoY=0$, while the TSHT procedure generally exhibits good direction detection performance in almost all scenarios except when $\beta_\YtoX=0$, its estimates have the largest RMSE and near-zero coverage probabilities for $\beta_\YtoX$. As for the Egger and IVW methods, they yield better estimation results for $\beta_\YtoX$, but their direction detection accuracy is lower compared to TSHT. In case (b) of $\beta_\YtoX=0$, the three existing methods exhibit poor direction detection and insufficient, sometimes even near-zero, coverage probabilities for estimating $\beta_\YtoX$.
In case (c) of $\beta_\XtoY=0.5$,  Egger consistently estimates the correct direction in almost all scenarios, and the TSHT and IVW methods behave similarly, except when $\beta_\YtoX=0$. However,  none of these existing methods consistently yield small RMSEs or reasonable coverage probabilities when estimating the two parameters across varying values of $\beta_\YtoX$. In case (d) of $\beta_\YtoX=0.5$,  the TSHT procedure performs well in detecting direction and estimating the parameter $\beta_\XtoY$. However, it exhibits disastrous performance in estimating $\beta_\YtoX$. On the other hand, the other two methods, Egger and IVW, perform better in estimating $\beta_\YtoX$, showing smaller RMSEs and higher coverage probabilities compared to TSHT. However, they suffer from worse performance in direction detection and estimation of $\beta_\XtoY$.

\section{Application to UK Biobank data}
\label{sec-data}
In this section, we apply the proposed method to infer the causal relationships among multiple traits with the UK Biobank data, a large prospective cohort study \citep{collins2012makes}. We are interested in exploring the causal relationships between body mass index (BMI) and two metabolic traits, including systolic blood pressure (SBP) and glycated hemoglobin A1c (HbA1c). This investigation is driven by substantial evidence suggesting that obesity is a significant risk factor for conditions such as hypertension and diabetes \citep{sullivan2008impact,franks2017causal,walsh2018trajectories,malone2019does}. Conversely, numerous studies have suggested that metabolic abnormalities can influence weight fluctuations \citep{Krzysztof2005,malone2019does}.

We  study the causal relationship between BMI and SBP in Section \ref{sec-data-1} and the relationship between BMI and HbA1c in Section \ref{sec-data-2}. For a pair of traits, we consider genetic variants which are associated with both traits as IVs, and a list of candidate IVs are selected based on the p-values of GWAS summary statistics obtained from the MRC-IEU consortium \citep{elsworth2020mrc}, and linkage disequilibrium of the genetic variants obtained from the EUR population of 1000 genomes data \citep{fairley2020international}. The significance threshold for $p$-value is set as $10^{-3}$ and LD-clumping threshold is set as $10^{-6}$ with R package ``TwoSampleMR'' \citep{twosamplemr}. 

In our study using the UK Biobank data, we include all participants aged 18 and above at the time of enrollment who were not taking any medications for lowering cholesterol, managing blood pressure, or treating diabetes, nor were they using exogenous hormones. This selection criteria ensures a more naturalistic assessment of the metabolic traits under investigation, free from pharmacological influences. We extract the genetic variants identified in the previous step, and exclude variants with observed minor allele frequencies lower than 5\%.
Details can be found in the supplemental material (Section 
S4.1).
To evaluate the robustness of the causal inference results, we also consider a different set of candidate IVs for each study  which is selected at a different significance level and the results are reported in Section 
S4.2
of the supplementary material.

\subsection{Causal relationship between body mass index and systolic blood pressure}
\label{sec-data-1}

We first apply the proposed method and existing methods to a well-studied pair of trait in MR literature, body mass index (BMI) and systolic blood pressure (SBP). 
The number of candidate IVs is 1553 and the sample size is 161988.
\begin{table}[!htbp]
	\centering
	\begin{tabular}{|c|c|c|c|c|c|}
		\hline
		Method    & $\widehat{\mathcal{H}}$ & $\hat\beta_\XtoY$ & $\widehat{\CI}_{\XtoY}(0.05)$ & $\hat\beta_\YtoX$& $\widehat{\CI}_{\YtoX}(0.05)$\\
		\hline
		PCH  & 1 & 0.352&$(0.281,0.422)$ & $-0.009$&$(-0.017,-0.001)$\\
		TSHT &1 &0.413 &(0.305, 0.520) &$-0.010$  &$(-0.018,-0.002)$\\
		Egger &  0&0.263 & $ (-0.032, 0.557) $&$-0.012$& $(-0.034, 0.010)$\\
		IVW &  2 &0.513 &$(0.339, 0.686)$ &~~0.041 & $(0.027, 0.055)$\\
		\hline
	\end{tabular}
	
	\vspace{2mm}
	\caption{Estimated causal direction and estimated confidence interval for the casual effects between traits BMI (X) and SBP (Y) based on the proposed (PCH) and existing methods (TSHT, Egger, and IVW).}
	\label{tab-bmi-sbp}
\end{table}
As in Table \ref{tab-bmi-sbp}, the proposed PCH estimates the causal relationship to be BMI $\rightarrow$ SBP and BMI has a positive effect on SBP. We see that the estimated confidence interval for $\beta_{\YtoX}$ does not cover zero at 95\% confidence level, which seems to be contradictory to $\widehat{\mathcal{H}}$. In fact, when estimating the causal direction, the confidence level in use is $1-1/n$ as in Algorithm \ref{alg:pch} and the corresponding confidence interval for $\beta_{\YtoX}$ covers zero. Hence, there is no contradiction in the results of PCH.
Th results of PCH agree with the findings based on many one-directional MR methods \citep{zhao2020statistical,spiller2022interaction} and our method convinces that there is no reverse causal effect. In comparison to Table 
S2
of the supplementary material, we see that the results of PCH remain robust when using a smaller and strong set of candidate IVs.

In view of the results of existing methods, TSHT has the same direction estimation result as PCH. Egger detects no significant causal direction but IVW detects bi-directional causal effects. The results of IVW are not robust when using a smaller set of candidate IVs. As we mentioned before, these methods have no guarantees when applied in two directions and therefore these results are less trustworthy.

\subsection{Causal relationship between body mass index and glycated hemoglobin}
\label{sec-data-2}

In this subsection, we apply the proposed method and existing methods to infer the causal relationship between BMI and Glycated hemoglobin (HbA1c). After prescreening, there are $n=147794$ samples and $p=672$ candidate IVs.

\begin{table}[!htbp]
	\centering
	\begin{tabular}{|c|c|c|c|c|c|}
		\hline
		Method    & $\widehat{\mathcal{H}}$ &$\hat\beta_\XtoY$& $\widehat{\CI}_{\XtoY}(0.05)$ & $\hat\beta_\YtoX$& $\widehat{\CI}_{\YtoX}(0.05)$\\
		\hline
		PCH  & 1 &0.206& $(0.179,0.234)$ &$-0.006$& $(-0.039, 0.023)$\\
		TSHT & 1 &0.161&(0.130, 0.191) & $-0.008$& $(-0.036, 0.021)$ \\
		Egger &  0&0.235&$ (0.101, 0.370) $&~~0.011& $(-0.069, 0.091)$\\
		IVW &  0 &0.127&$(0.049, 0.206)$ &~~0.116& $(0.045, 0.188)$\\
		\hline
	\end{tabular}
	
	\vspace{2mm}
	\caption{Estimated causal direction and estimated confidence interval for the casual effects between traits BMI (X) and HbA1c (Y) based on the proposed (PCH) and existing methods (TSHT, Egger, and IVW).}
	\label{tab-bmi-hba}
\end{table}

We see from Table \ref{tab-bmi-hba} that the proposed method detects the causal direction as BMI$\rightarrow$HbA1c and BMI has a positive effect on HbA1c. The TSHT method has the same detection result but IVW and Egger detects no significant causal relationship between BMI and HbA1c.  \citet{hu2020exploring} also finds positive causal relationship between BMI and HbA1c through one-directional MR analysis.

\section{Discussion}\label{sec:discussion}

In this paper, we have proposed a method to make inference of causal directions and effects between two traits with possibly
bi-directional relationships and unmeasured confounding. We provide two 
identification approaches. The first approach relies on the plurality rule holding for one known direction and the CH condition for the other. The second approach addresses the more realistic scenario where the direction in which the plurality rule holds is unknown, establishing the identification results under an enhanced plurality rule and CH conditions.
Then we develop the PCH method to infer the
causal relationships and make inference on the causal effects.  
We show that the proposed method can consistently  estimate the true causal direction and provides valid confidence intervals for the causal effects.

This paper may be extended or improved in several directions. Firstly, the identification and inference framework proposed here primarily focuses on low dimensions with invalid instruments. An interesting avenue for future work is to extend this framework to high-dimensional settings, allowing the dimension of instruments to potentially exceed the sample size.
Secondly, the proposed framework can be improved by considering more flexible models, such as nonlinear ones, to better capture the relationships between the traits and instruments.
Thirdly, it is of interest to extend our results to the task of learning causal networks among multiple traits, especially those with feedback loops and unmeasured confounding.
We plan to pursue these and other related issues in future research.

	\bibliographystyle{apalike}
\bibliography{mybib}

\end{document}